# Quantum scattering theory of Fock states in high-dimensional spaces

Jingfeng Liu[1,2], Ming Zhou[2], Zongfu Yu[2*]

1. College of Electronic Engineering, South China Agricultural University, Guangzhou 510642, China
2. Department of Electrical and Computer Engineering, University of Wisconsin – Madison, 53706, U.S.A.

* zyu54@wisc.edu

**Abstract**: A quantum scattering theory is developed for Fock states scattered by two-level systems in the free space. Compared to existing scattering theories that treat incident light semi-classically, the theory fully quantizes the incident light as Fock states. This non-perturbative method provides exact scattering matrix.

The resonant scattering of photons by a two-level system (TLS) is one of the most fundamental processes in light-matter interactions. In traditional experiments, this scattering process was often studied using classical incident light. Approximated theories that treat incident photons as semi-classical fields are often adequate for explaining the experimental observation. In the past decades, the progress of experimental techniques has allowed us to prepare incident light in Fock states. This non-classical light has quickly gained technological importance because of the applications in quantum information. Meanwhile, a new theoretical challenge has emerged: how to model the quantum scattering of Fock states, particularly for multi-photon states. Here in the letter, we propose a theoretical framework to address this challenge and develop a quantum scattering theory for Fock states in three-dimensional spaces.

The Hamiltonian for a TLS in the free space has a rather simple form:

$$H = \hbar\omega_e \sigma^\dagger \sigma + \sum_k \hbar\omega_k a_k^\dagger a_k + \sum_k i\hbar g_k(a_k^\dagger \sigma - a_k \sigma^\dagger) \qquad (1)$$

The first and the second terms are the free Hamiltonian of a two-level system and free-space photons, respectively. The third term describes the interaction between them under the dipole and rotating-wave approximations. Here $\hbar$ is reduced Planck constant and $i = \sqrt{-1}$. $\omega_e$ is the transition frequency of the TLS. $\sigma^\dagger$ and $\sigma$ are the raising and lowering operator, respectively. $\omega_k$ and $\boldsymbol{k}$ are the angular frequency and the wavevector of photons, respectively. $a_k^\dagger$ and $a_k$ are the bosonic creation and annihilation operator of photons, respectively. The coupling coefficient is $g_k = \boldsymbol{d} \cdot \boldsymbol{e_k}\sqrt{\omega_e/(2\hbar\varepsilon_0 L^3)}$ where the transition dipole moment is $\boldsymbol{d}$ and $\boldsymbol{e_k}$ is the polarization of the light. $L^3$ is the normalization volume.

Our goal is to solve the scattering problem governed by Eq. (1). Many approximated approaches have been developed. The common issue with these methods is that they do not fully quantize the incident light. For example, a widely used method is based on the master equations[1]. The incident field is treated as a *c*-number $\boldsymbol{E}_{in}$ that drives the polarization of the TLS in the master equation. The scattering field is calculated based on the semi-classical radiation of the polarized TLS. This method can be useful under the right conditions. For instance, in the weak-light limit $\boldsymbol{E}_{in} \to 0$, it correctly predicts the scattering of single photons. However, this method completely fails for multi-photon scattering. More importantly, the semi-classical nature leaves our understanding on a less satisfying foundation.

In this Letter, we develop a quantum scattering theory. In particular, we fully quantize the incident photons as Fock states. It is a non-perturbative scattering theory that solves for the eigenstates of the Hamiltonian in Eq. (1) without any approximation. The theoretical framework is built upon methods developed for electrons scattered by magnetic impurities in condensed matter physics[2,3] as well as recent progress in one-dimensional waveguide quantum electrodynamics[4-19]. Although we illustrate the theory based on single photons, the framework is fundamentally compatible with and can be extended to multi-photons.

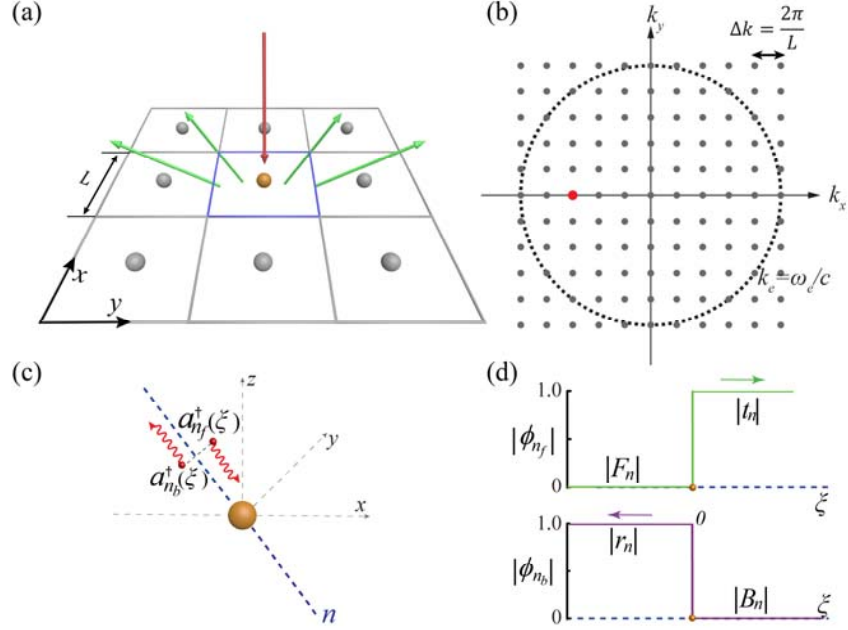

Fig. 1 (color online). (a) Periodic boundary conditions are set up for a single TLS in the free space. $L$ is the periodicity and we will take $L \to \infty$ at the end of the derivation. (b) Distribution of channels in the $k$-space. Due to the periodicity, scattered light has discrete wavevectors in the $\boldsymbol{k}_{xy}$-plane, represented by dots. (c) Spatial operator $a^\dagger_{n_{f/b}}(\xi)$ creates a forward or backward moving photon at the location $\xi$ in the $n^{th}$ channel. (d) Schematic of the magnitude of the spatial wave-function $\left|\phi_{n_{f/b}}(\xi)\right|$ for the forward(green line) and backward(purple line) directions in the $n^{th}$ channel.

We start by first discretizing the 3D continuum. It is realized by using a periodic boundary condition in the $x$-$y$ plane (Fig. 1a). The period is $L$. At the end of the derivation, we will take the limit of $L \to \infty$ to recover the case of a single and isolated TLS in the free space. Because of the periodicity, a normally incident photon can only be scattered to a set of discrete directions. These directions are defined by the waves' in-plane wavevectors $\boldsymbol{k}_{xy} = (m_x, m_y)2\pi/L$, where $m_{x,y}$ are integers (Fig. 1b). We call these directions *channels*. As a convention of the notation, the channel $(m_x, m_y)$ in the upper semi-infinite space also includes the waves in lower semi-infinite space in the direction of $(-m_x, -m_y)2\pi/L$. Channels are all located within the circle of $k_e = \omega_e/c$ for the interested frequency range around the resonant frequency $\omega_e$. The total number is $N = \pi \lfloor L/\lambda_e \rfloor^2$, where $\lambda_e = 2\pi/k_e$ is the resonant wavelength. The floor operator $\lfloor A \rfloor$ gives the largest integer smaller than $A$. There are also two propagation directions in each channel, which are labeled with subscripts $f$ (forward) and $b$ (backward), respectively. There are also two polarizations in each channel. Using channels, we can convert the Hamiltonian to:

$$H = \hbar\omega_e \sigma^\dagger \sigma + \sum_{n=1}^{N}\sum_k \hbar\omega_{k,n}\left(a^\dagger_{k,n_f} a_{k,n_f} + a^\dagger_{k,n_b} a_{k,n_b}\right)$$
$$+ \sum_{n=1}^{N}\sum_k i\hbar \mathbb{N}(\theta_n, \varphi_n)\, g_k \left[\left(a^\dagger_{k,n_f} + a^\dagger_{k,n_b}\right)\sigma - (a_{k,n_f} + a_{k,n_b})\sigma^\dagger\right], \quad (2)$$

It is important to note that the wavenumber $k$ now is a scalar because the information of the propagation direction is absorbed into the channel definition. Because of periodic boundary condition, we need to normalize the coupling coefficient with a factor $\mathbb{N}(\theta_n, \varphi_n) = \sqrt{(cos^2\varphi_n - sin^2\varphi_n)/cos\theta_n}$, where $\theta_n$ and $\varphi_n$ are the polar and azimuthal angles of the $n^{th}$ channel, respectively[20,21].

In order to solve for the spatial wavefunctions, we further convert the Hamiltonian to a real-space representation by applying the following Fourier transformation:

$$a^\dagger_{k,n_{f/b}} = \frac{1}{\sqrt{L}}\int_{-\infty}^{\infty} d\xi\, a^\dagger_{n_{f/b}}(\xi)\exp(ik\xi), \quad (3a)$$

$$a_{k,n_{f/b}} = \frac{1}{\sqrt{L}}\int_{-\infty}^{\infty} d\xi\, a_{n_{f/b}}(\xi)\exp(-ik\xi), \quad (3b)$$

where $\xi$ is the spatial coordinate alone the channel $n$ as shown in Fig. 1c. The operator $a^\dagger_{n_{f/b}}(\xi)$ creates a forward or backward moving photon at location $\xi$ in the $n^{th}$ channel. This transformation has been used for the scattering theory in one-dimensional continuum such as waveguide QED[24-26]. Here we extend the method to the three-dimensional space. Substituting Eq. (3) into Eq. (2), we get the real-space Hamiltonian

$$H = \hbar\omega_e \sigma^\dagger \sigma + \sum_{n=1}^{N} \int_{-\infty}^{+\infty} d\xi \left\{ \left[ (-i\hbar c) a_{n_f}^\dagger(\xi) \frac{d}{d\xi} a_{n_f}(\xi) \right. \right.$$
$$\left. + (i\hbar c) a_{n_b}^\dagger(\xi) \frac{d}{d\xi} a_{n_b}(\xi) \right] \quad (4)$$
$$\left. + i\hbar g_n \delta(\xi) \{ [a_{n_f}^\dagger(\xi) + a_{n_b}^\dagger(\xi)]\sigma - [a_{n_f}(\xi) + a_{n_b}(\xi)]\sigma^\dagger \} \right\},$$

where $c$ is the speed of light and $g_n = \sqrt{\frac{L(\cos^2\varphi_n - \sin^2\varphi_n)}{\cos\theta_n}} \, g_k$. $\delta(\xi)$ is the Dirac delta function.

For a single photon, the eigenstate of the Hamiltonian in Eq. 4 can be written as

$$|\phi\rangle = \left( \sum_{n=1}^{N} \int d\xi \left( \phi_{n_f}(\xi) a_{n_f}^\dagger(\xi) + \phi_{n_b}(\xi) a_{n_b}^\dagger(\xi) \right) + e\sigma^\dagger \right) |0, g\rangle, \quad (5)$$

where $|0, g\rangle$ represents the ground state. $e$ is the probability amplitude of the atom in the excited state; $\phi_{n_{f/b}}(\xi)$ are the spatial distribution of the amplitudes in channels.

We can now directly evaluate $e$ and $\phi_{n_{f/b}}(\xi)$ using the time-independent Schrödinger equation $H|\phi\rangle = \hbar\omega|\phi\rangle$ and obtain

$$-ic \frac{d}{d\xi} \phi_{n_f}(\xi) + ig_n \delta(\xi) e = \omega \phi_{n_f}(\xi), \quad (6a)$$

$$ic \frac{d}{d\xi} \phi_{n_b}(\xi) + ig_n \delta(\xi) e = \omega \phi_{n_b}(\xi), \quad (6b)$$

$$\omega_e e - \sum_{n=1}^{N} ig_n \left( \phi_{n_f}(0) + \phi_{n_b}(0) \right) = \omega e, \quad (6c)$$

These linear differential equations can be easily solved. Specifically, Eq. (6a) reduces to $-ic \frac{d}{d\xi} \phi_{n_f}(\xi) = \omega \phi_{n_f}(\xi)$ for $\xi \neq 0$, which has a simple solution

$$\phi_{n_f}(\xi) = F_n e^{ik\xi} \theta(-\xi) + t_n e^{ik\xi} \theta(\xi), \quad (7a)$$

where the wave number $k = \omega/c$. Similarly for the backward directions, Eq. (6b) leads to

$$\phi_{n_b}(\xi) = r_n e^{-ik\xi} \theta(-\xi) + B_n e^{-ik\xi} \theta(\xi). \quad (7b)$$

In a scattering process, coefficients $F_n$ and $B_n$ can be interpreted as the incident amplitudes of the photon in the forward and backward directions, respectively. $t_n$ and $r_n$ represent the transmitted amplitudes in the forward and backward directions, respectively. These coefficients must satisfy the boundary condition at the location of the atom: $F_n + t_n = r_n + B_n$. The boundary conditions at infinity are determined by the incident condition.

Now we consider a single photon incident in the direction of the $l^{th}$ channel in the forward direction. Then we have $F_n = \delta_{nl}$ and $B_n = 0$. The wavefunctions $|\phi_{n_{f/b}}(\xi)|$ are schematically shown in Fig. 1d for a channel $n \neq l$. The forward and backward scattering amplitudes are represented by $t_n$ and $r_n$, respectively. Using Eqs. 6-7, we can calculate these coefficients as

$$t_n = \frac{-ig_n g_l/c}{(\omega - \omega_e) + i\Gamma_0/2} + \delta_{nl}, \tag{8a}$$

$$r_n = \frac{-ig_n g_l/c}{(\omega - \omega_e) + i\Gamma_0/2}, \tag{8b}$$

$$e = \frac{-ig_l}{(\omega - \omega_e) + i\Gamma_0/2}, \tag{9}$$

where $\Gamma_0 = \frac{d^2 \omega_e^3}{3\pi\hbar\varepsilon_0 c^3}$ is the spontaneous emission rate of the TLS in free space (see appendix B for its derivation).

Now we have the complete spatial wavefunction of the eigenstates, from which we can obtain all the characteristics of the scattering process. We will briefly discuss a few examples below.

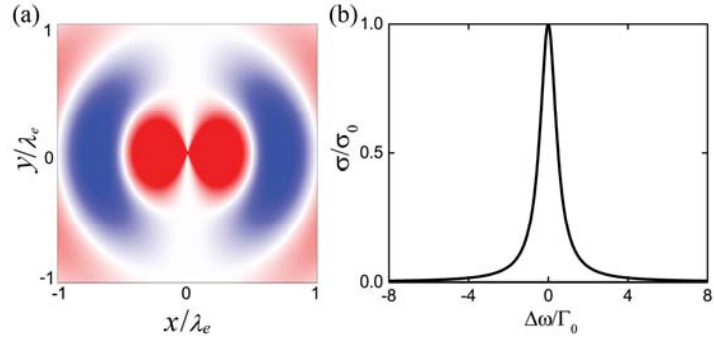

Fig.2 (a) The real part pattern of the scattered amplitude $\phi_p(\mathbf{r})$ for a single photon scattered by a TLS (b) The spectrum of the scattering cross section.

**Scattered fields**. The spatial distribution of the scattered photon can be directly evaluated by summing the amplitudes in all channels $\phi_p(\mathbf{r}) = \sum_{n=1}^{N} [\phi_{n_f}(\xi_n) + \phi_{n_b}(\xi_n)]$. $\xi_n$ is calculated by projecting the position $\mathbf{r}$ onto the $n^{th}$ channels. Specifically, we consider a TLS with a dipole moment induced by a linearly-polarized incident light. The incident photon propagates along $x$ axis and is polarized alone the $z$ direction (into the plane in Fig. 2a). Fig. 2a shows the real part of the scattering wavefunction $\phi_p(\mathbf{r})$ in the $xz$ plane. It clearly shows a dipole radiation profile.

**Cross section**. We can also easily calculate the differential and total cross sections[27,28]. For example, the total scattering cross section is

$$\sigma(\omega) = \frac{\sum_{n=1}^{N}(t_n^\dagger)t_n + \sum_{n=1}^{N}(r_n^\dagger)r_n}{1/L^2} \tag{10}$$

Substituting $r_n$ and $t_n$ into Eq. (10), we obtain

$$\sigma(\omega) = \frac{3\lambda_e^2}{2\pi}\frac{(\Gamma_0/2)^2}{(\omega-\omega_e)^2+(\Gamma_0/2)^2} \tag{11}$$

The cross section spectrum is shown in Fig. 2b, which shows the typical Lorentzian lineshape with a bandwidth defined by the spontaneous emission rate $\Gamma_0$.

**Winger time delay.** The complete wavefunction also allows us to calculate the group delay $\tau = \frac{d\varphi}{d\omega}$ for the photon when scattered by a TLS. The scattering phase $\varphi$ can be directly evaluated from the wave function as $\varphi(\omega) = \arctan\{-\frac{\Gamma_0}{[2(\omega-\omega_e)]}\}$, which leads to a Winger time-delay $\tau = \frac{\Gamma_0/2}{(\omega-\omega_e)^2+(\Gamma_0/2)^2}$.

All these results confirm the calculation based on semi-classical scattering theories[29-32]. Here below, we further show that the theory can easily accommodate multiple TLS scatterers. Specifically, we use two TLSs as an example to illustrate the method. The Hamiltonian can be written as

$$\begin{aligned} H_0 = & \sum_{m=1}^{2} \hbar\omega_m \sigma_m^\dagger \sigma_m + \hbar\Omega_{12}(\sigma_1^\dagger\sigma_2 + \sigma_2^\dagger\sigma_1) \\ & + \sum_k \hbar\omega_k a_k^\dagger a_k + \sum_{m=1}^{2}\sum_k i\hbar g_{mk}(a_k^\dagger \sigma_m - a_k \sigma_m^\dagger), \end{aligned} \tag{12}$$

where $\Omega_{12}$ is the strength of induced dipole-dipole interaction between the two TLSs[1,33]. Following similar procedures, we convert it to a real-space representation using channels:

$$\begin{aligned} H = & \sum_{m=1}^{2} \hbar\omega_m \sigma_m^\dagger \sigma_m + \hbar\Omega_{12}(\sigma_1^\dagger\sigma_2 + \sigma_2^\dagger\sigma_1) \\ & + \sum_{n=1}^{N}\int_{-\infty}^{\infty} d\xi \left\{(-i\hbar c)a_{n_f}^\dagger(\xi)\frac{d}{d\xi}a_{n_f}(\xi)\right. \\ & \left. + (i\hbar c)a_{n_b}^\dagger(\xi)\frac{d}{d\xi}a_{n_b}(\xi)\right\} \\ & + i\hbar\sum_{n=1}^{N}\int_{-\infty}^{\infty}d\xi \sum_{m=1}^{2} g_{n,m}\delta(\xi-\xi_{n,m})\{[a_{n_f}^\dagger(\xi)+a_{n_b}^\dagger(\xi)]\sigma_m \\ & - [a_{n_f}(\xi)+a_{n_b}(\xi)]\sigma_m^\dagger\}. \end{aligned} \tag{13}$$

where $\xi_{n,m}$ is the projected location of the $m^{\text{th}}$ TLS in the $n^{\text{th}}$ channel. The general form of the eigenfunction for a single excitation can be written as

$$|\phi\rangle = \sum_{n=1}^{N}\int d\xi \left(\phi_{n_f}(\xi)a^{\dagger}_{n_f}(\xi) + \phi_{n_b}(\xi)a^{\dagger}_{n_b}(\xi)\right)|0, g_1, g_2\rangle \tag{14}$$
$$+ e_1\sigma_1^{\dagger}|0, g_1, g_2\rangle + e_2\sigma_2^{\dagger}|0, g_1, g_2\rangle,$$

where $|0, g_1, g_2\rangle$ is the ground state.

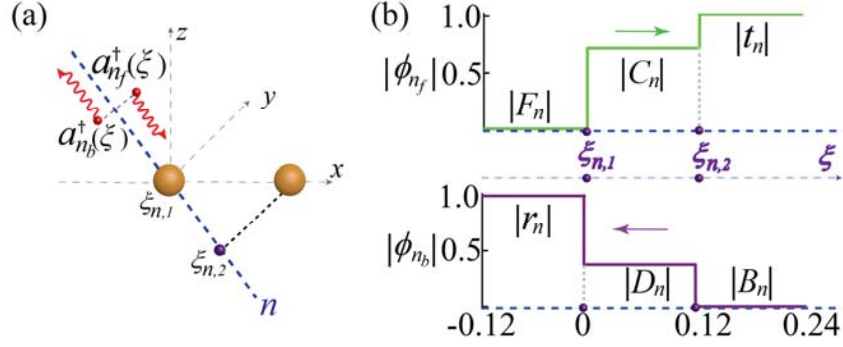

Fig.3 (a) Schematic of a channel with two TLSs. Single the first TLS is located at the origin, $\xi_{n,1} = 0$ for all channels. (b) Schematic of the magnitude of the photon wavefunction $|\phi_{n_{f/b}}(\xi)|$ for the forward(green line) and backward(purple line) directions in $n^{th}$ channel at the resonant excitation.

The forward and backward wavefunctions have three distinct segments as divided by two TLSs:
$$\phi_{n_f}(\xi) = e^{ik\xi}\left[F_n\theta(\xi_{n,1} - \xi) + C_n\theta(\xi - \xi_{n,1})\theta(\xi_{n,2} - \xi) + t_n\theta(\xi - \xi_{n,2})\right], \tag{15a}$$
$$\phi_{n_b}(\xi) = e^{-ik\xi}\left[r_n\theta(\xi_{n,1} - \xi) + D_n\theta(\xi - \xi_{n,1})\theta(\xi_{n,2} - \xi) + B_n\theta(\xi - \xi_{n,2})\right]. \tag{15b}$$

All the coefficients $F_n, C_n, t_n, r_n, D_n$ and $B_n$ can be obtained by solving the Schrödinger equation. They are also constraint by the boundary conditions at the locations of the two TLSs and at infinity.

Similar to the single TLS case, $F_n$ and $B_n$ represent the incident photon in the forward and backward directions, respectively. $F_n = B_n = 0$ for all channels except for the incident directions, as shown schematically in Fig. 3b. $t_n$ and $r_n$ are the amplitudes of the scattered photon in the forward and backward directions, respectively.

Different from the single TLS case, here we have additional terms, represented by $C_n$ and $D_n$. They are the waves in-between the two TLSs. These waves induce the radiative interactions among the TLSs. They are responsible for collective effects, such as the superradiant spontaneous emission[34-38].

We now consider a specific example of a single photon scattered by two identical TLSs located at $x = 0$ and at $x = 0.15\lambda_e$. The directions of their dipole moments are induced by the incident light.

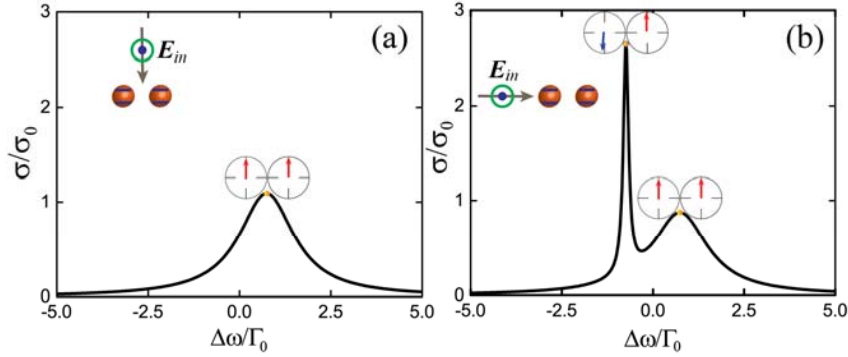

Fig. 4 The scattering cross section of the photon incident from (a) the normal direction and (b) the axial direction. The complex amplitudes of $e_1$ and $e_2$ of two TLSs are illustrated by vectors in a circle as shown in insets. Two red arrows in (a) show that the two TLS are in-phase at the peak frequency. On the other hand, the sharp peak in (b) exhibit opposite phases for the amplitudes of the two TLSs.

Figure 4 shows the spectra of the cross section calculated from the quantum scattering theory. For a single photon incident from the normal direction (Fig. 4a), the spectral bandwidth is nearly doubled compared to the case of single TLS (Fig. 2b). This bandwidth broadening is the manifest of the superradiance in the scattering process. The superradiance can be more clearly observed by examining the complex amplitudes $e_1$ and $e_2$ of the two TLSs, which are shown as the inset of Fig. 4a. They show the same phase and amplitude.

When a single photon is incident from the side in the axial direction (Fig. 4b), a second sharp peak appears in the spectrum of the cross section. This peak is associated with the sub-radiant oscillation of the two TLSs. The excitation amplitudes of the two TLSs exhibit opposite phases as shown by the inset of Fig. 4b. The opposite-phase oscillation strongly reduces the radiation rate, resulting in an extremely narrow linewidth. It also leads to a larger cross section $\sigma = 2.5\sigma_0$ than the linear addition for two atoms, i.e. $2\sigma_0$. Since the two TLSs have opposite phases, this oscillation mode cannot be excited from the normal direction and thus is absence in spectrum shown in Fig. 4a.

In conclusion, we have developed a quantum scattering theory for the quantum transport of Fock states in the free space. Using the theory, we show that two TLSs can lead to both superradiant and subradiant scattering effects. We have illustrated our theory based on single photons. We emphasize that the scattering properties of single photons can be derived without fully quantizing the incident light. Our exact theory confirms those results obtained in semi-classical treatment. Most importantly, our theoretical framework is fundamentally compatible and can be extended to multi-photon case, which will be discussed in another paper.

**Acknowledgement:**
The authors acknowledge the financial support by NSF Grant ECCS 1405201, National Natural Science Foundation of China (Grant 1204089, Grant 11334015) and China Scholarship Council.

## Appendix A: The Spontaneous Emission Rate $\Gamma_0$

Here we solve the Schrodinger equation and show the derivation of Eqs. (8a-9). We directly substitute the general form of the wavefunction $\phi_{n_f}(\xi) = e^{ik\xi}[\theta(-\xi)\delta_{nl} + t_n\theta(\xi)]$ and $\phi_{n_b}(\xi) = e^{-ik\xi}[r_n\theta(-\xi)]$ into Eq. (6a) and (6b) and obtain

$$t_n = \frac{g_n}{c}e + \delta_{nl}, \tag{A1}$$

$$r_n = \frac{g_n}{c}e. \tag{A2}$$

Based on the definition $\theta(0) = 1/2$ of the step function, we can write $\phi_{n_f}(0) = [\delta_{nl}/2 + t_n/2]$ and $\phi_{n_b}(0) = r_n/2$. Substituting $\phi_{n_f}(0)$ and $\phi_{n_b}(0)$ into Eq. (6c), we arrive at

$$\omega_e e - \sum_n^N ig_n\left(\frac{\delta_{nl} + t_n}{2} + \frac{r_n}{2}\right) = \omega e, \tag{A3}$$

Combining Eqs. (A1-A3), we obtain the amplitude of the excited state

$$e = \frac{-ig_l}{(\omega - \omega_e) + i\sum_n^N(g_n^2/c)}. \tag{A4}$$

Here we could define the spontaneous emission rate as

$$\Gamma_0 = 2\sum_n^N \frac{g_n^2}{c}, \tag{A5}$$

$g_n$ is given in Eq. (4).

Next we calculate $\Gamma_0$. The summation in Eq. (A5) is carried over all $N$ channels as shown in Fig. 2. First, we take care of the summation of the two polarization directions $\boldsymbol{e}_{n1}$ and $\boldsymbol{e}_{n2}$ as shown in Fig. (A1).

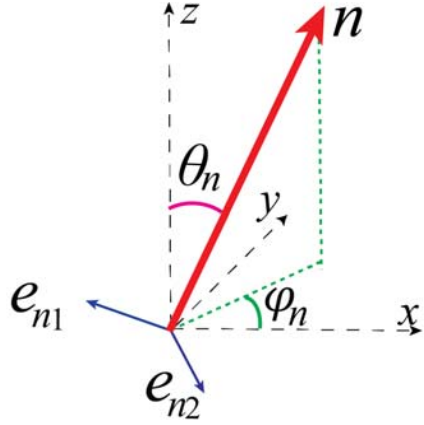

Fig. A1 (color online). The polarization direction of the dipole moment $\boldsymbol{d}$ is alone the $z$-axis. $\boldsymbol{e}_{n1}$ and $\boldsymbol{e}_{n2}$ are the two polarizations. $\boldsymbol{e}_{n1}$ and $\boldsymbol{e}_{n2}$ are orthogonal to each other, and both are perpendicular the $n^{th}$ channel.

$$g_n^2 = \frac{cos^2\varphi_n - sin^2\varphi_n}{cos\theta_n} \frac{\omega_e}{2\hbar\varepsilon_0 L^2} \sum_{s=1}^{2}|\boldsymbol{d}\cdot \boldsymbol{e}_{ns}|^2$$
$$= \frac{d^2\omega_e}{2\hbar\varepsilon_0 L^2}\frac{cos^2\varphi_n - sin^2\varphi_n}{cos\theta_n}(cos^2\alpha_n + cos^2\beta_n), \quad (A6)$$

where $\alpha_n$ and $\beta_n$ are the polar angles of $e_{n1}$ and $e_{n2}$, respectively. Without loss of generality, here we define the dipole directions along the $z$-axis directions. $\theta_n$ and $\varphi_n$ are the polar and azimuthal angles of the $n^{th}$ channel. Using the law of cosines $cos^2\alpha_n + cos^2\beta_n = 1 - cos^2\theta_n$, we can further simplify Eq. A6 to

$$g_n^2 = \frac{d^2\omega_e}{2\hbar\varepsilon_0 L^2}\frac{cos^2\varphi_n - sin^2\varphi_n}{cos\theta_n}(1 - cos^2\theta_n). \quad (A7)$$

Substituting Eqs. (A7) into Eq. (A5), we have

$$\Gamma_0 = \sum_{l=1}^{N}\frac{2g_0^2}{c}\frac{cos^2\varphi_n - sin^2\varphi_n}{cos\theta_n}(1-cos^2\theta_n)$$
$$= \frac{L^2 g_0^2}{\lambda_e^2 c}\sum_{l=1}^{2*N}\frac{cos\varphi_n + sin\varphi_n}{L/\lambda_e cos\theta_n}\frac{cos\varphi_n - sin\varphi_n}{L/\lambda_e sin\theta_n}sin^3\theta_n, \quad (A8)$$

where $g_0^2 = d^2\omega_e/(2\hbar\varepsilon_0 L^2)$. When $L \to \infty$, we can convert the summation in Eq. (A8) to the integral form using the transformation: $d\theta_n = (cos\varphi_n + sin\varphi_n)/(L/\lambda_e cos\theta_n)$, $d\varphi_n = (cos\varphi_n - sin\varphi_n)/(L/\lambda_e sin\theta_n)$.

$$\Gamma_0 = \frac{L^2 g_0^2}{\lambda_e^2 c}\int_0^\pi\int_0^{2\pi} sin^3\theta_n\, d\theta_n d\varphi_n = \frac{d^2\omega_e^3}{3\pi\hbar\varepsilon_0 c^3} \quad (A9)$$